\documentstyle[preprint,aps]{revtex}

\begin{document}
\title{Energy Dependence of Quasi-Particle Relaxation in a Disordered Fermi Liquid}
\author{T. Schmidt$^1$\cite{address}, P. K\"onig$^1$, E. McCann$^2$, Vladimir I. Fal'ko$^2$, and
R. J. Haug$^{1}$}
\address{$^1$ Institut f\"ur Festk\"orperphysik, Universit\"at
Hannover, Appelstr. 2, 30167 Hannover, Germany\\
$^{2}$ Physics Department, Lancaster University, LA1 4YB 
Lancaster, United Kingdom}
\date{8.11.00}
\maketitle

\begin{abstract}
A spectroscopic method is applied to measure the inelastic quasi-particle
relaxation rate in a disordered Fermi liquid. The quasi-particle relaxation
rate, $\gamma $ is deduced from the magnitude of fluctuations in the local
density of states, which are probed using resonant tunneling through a
localized impurity state. We study its dependence on the excitation
energy $E$ measured from the Fermi level. In a disordered metal (heavily
doped GaAs) we find that $\gamma \propto E^{3/2}$ within the experimentally
accessible energy interval, in agreement with the Altshuler-Aronov theory
for electron-electron interactions in diffusive conductors.
\end{abstract}


The quasi-particle description of electrons represents a common approach to
understanding kinetic and thermodynamic phenomena in metals. It relies on a
certain stability of quasiparticle excitations in a many-body system, which
requires that the quasi-particle excitation energy, $E$, determined with
respect to the Fermi level, $E_{F}$, exceeds the broadening of such a
single-particle state, $\hbar \gamma (E)$, due to electron-electron
interactions.

In ideally pure metals, the relaxation rate, $\gamma ,$ of ballistic
quasi-particles at low energies is kept low by a diminishing of the phase
space available for inelastic electron-electron collision processes \cite
{pin66}. The phase space argument results in $\gamma \sim E^{2}$ dependence,
which has been confirmed by electron emission spectroscopy measurements
which probe electron states at energies $E\gg \hbar /\tau $, where $\tau $
is the elastic mean free path time due to residual impurities.
Quasi-particle decay at excitation energies $E<\hbar /\tau $ is accelerated
by the presence of disorder (which makes the electron states chaotically
random at  length scales longer than the elastic mean free path, $l=\tau
v_{F}$). The theory of interaction effects in diffusive media \cite{alt85}
predicts that, at zero temperature, the decay of quasi-particle excitations
with $E\ll \hbar /\tau $ in a disordered metal or heavily doped
semiconductor is slow enough to guarantee the existence of properly defined
quasi-particles. In particular, in bulk three-dimensional (d=3) and
two-dimensional (d=2) conductors, the quasi-particle relaxation rate is
expected to obey a power-law dependence, 
\begin{equation}
\gamma (E)=aE^{d/2}E_{F}^{(1-d/2)}(\lambda _{F}/l)^{d/2}.  \label{eqgamma}
\end{equation}

Despite the fundamental interest and clear theoretical predictions \cite
{alt85}, there are very few direct experimental measurements of the
quasi-particle decay time in disordered metals at small excitation energies $%
\varepsilon <\hbar /\tau $. Information about the inelastic decay of
non-equilibrium quasi-particles in dirty metals is often extracted from
energy relaxation rates in the electron thermalization process \cite{pot97}, 
$\gamma _{E}$. Otherwise, one studies the temperature dependence of a
dephasing rate, $\gamma _{\varphi }(T)$, of coherent carriers (determined
using weak localisation or universal conductance fluctuations analysis),
which can be treated as a measure of the efficiency of the interactions of
an electron with energy $E\leq T$ with equilibrium fluctuations of charges
produced by other electrons thermally distributed near the Fermi level in a
diffusive conductor \cite{imr97}. At high temperatures, both phase and
energy relaxation experiments show a certain agreement with theoretical
estimations \cite{ber84,was92,mar92,hub98}. However, recently reported data 
\cite{moh97,pot97} on two dynamical parameters mentioned above, $\gamma
_{E}(E)$ and $\gamma _{\varphi }(T)$, in Au wires and films, and also in
semiconductor heterostructures \cite{hub98} have indicated a certain
disagreement between theoretically predicted and experimentally observed
values of these two quantities, which has refuelled both theoretical and
experimental interest in the problem of quasi-particle lifetimes in a
disordered metal \cite{alt98}.

In the present paper, we report the results of a direct measurement of the
energy dependence of the inelastic decay rate $\gamma (E)$ of a
quasi-particle state in a disordered conductor. This study employs the
method of resonant tunneling spectroscopy using a discrete localised state
in a double-barrier structure, which has been applied earlier by A.Geim {\it %
et al} \cite{des94} to study 2D electrons in a heterostucture and by U.Sivan 
{\it et al} \cite{Sivan} to investigate the discrete spectra of quantum
pillars. It has been shown previously \cite{sch96,fal97,sch97,hol00} that,
by measuring the current-voltage (IV) characteristics and by deriving the
differential conductance in a system where the current passes through a
single resonant impurity state in the barrier, one can study features of the
single-particle spectrum of a disordered metal (playing the role of an
emitter). In a bulk material with a continuous spectrum, individual chaotic
quantum states formed by the interference of elastically scattered electrons
produce an effect known as fluctuations of the local density of states
(LDOS) \cite{ler92}. It consists of a random\ and coordinate-specific energy
dependence of the local density of states in a diffusive metal, $\nu (E)$ with a correlation
energy limited by inelastic broadening of quasi-particle states, $%
\hbar \gamma $. We observe a random pattern in $\nu (E)$ by sweeping a
single resonant impurity level against the electron spectrum in the emitter
within a finite range of excitation energies for a quasi-hole (an empty
state below $E_{F}$ in the emitter) left behind by the tunneled electron.
When the energetic width of an impurity level used in this process, $\Gamma $%
, is smaller than the inelastic broadening of single-particle levels in the
emitting electrode, $\hbar \gamma $, one can extract the latter
characteristic from the analysis of the amplitude of the LDOS fluctuations
pattern and its auto-correlation parameters. This has enabled us to measure
directly the energy dependence of the inelastic relaxation rate, $\gamma (E)$%
, in bulk degenerate heavily doped GaAs at low temperatures, which we find
to agree with the Altshuler-Aronov theory \cite{alt85,siv94} predicting $%
\hbar \gamma (E)=a\left( E^{3/2}/\sqrt{E_{F}}\right) \left( \lambda
_{F}/l\right) ^{3/2}$ for a 3D system. [Note, that under the condition of $%
\hbar \gamma (E)\ll E$ provided in Eq. (\ref{eqgamma}) for $\lambda _{F}\ll l
$, the inelastic decay rate $\gamma $ of a non-equilibrium quasi-particle
coincides with its decoherence rate.]

As mentioned above, the LDOS can be measured via resonant tunneling
through an impurity in a strongly asymmetric double-barrier heterostructure. 
Our microstructure consists of a 10 nm wide
GaAs quantum well and 5 and 8 nm wide Al$_{0.3}$Ga$_{0.7}$As barriers
sandwiched between doped GaAs contact layers with a donor concentration of $%
3.3\times 10^{17}$ cm$^{-3}$, as sketched in Fig. \ref{gv}(a). From this
material we fabricated a 2 $\mu $m diameter mesa, as depicted in the
scanning electron micrograph of Fig. \ref{gv}(b). The mesa contains a small
number of residual impurities in the nominally undoped quantum well. The
energetically lowest impurity state $S$ in the well is used as a
spectrometer for the LDOS imaging in the metallic emitter adjacent to the
thicker barrier as illustrated in Fig. \ref{gv}(c). At zero bias, $S$ lies
above the Fermi level in the emitter and is not available for resonant
transport, resulting in $I=0$ and $G=0$. This measurement has been performed
at the temperature $T=$20mK. Upon applying a finite bias voltage $V$, the
energetic position of a spectrometer S is shifted down to the energy $%
E=\alpha e(V-V_{S})$ below the Fermi level in the emitter, where the
prefactor $\alpha =0.50$ accounts for the fact that only part of the voltage
drops between emitter and spectrometer \cite{alpha}. When S crosses $E_{F}$
from above (at $V_{S}=9.8$mV), the current acquires a finite value
(plateaux) limited by the left, less transparent barrier and proportional to
the LDOS of occupied states in the emitter at the spectrometer position, $%
I\propto \nu $ \cite{sch96,fal97}. At higher bias voltages ($V_{H1}=14.6$mV
and $V_{H2}=15.5$ mV), other impurities (or, maybe, other excited states
from the same impurity) become involved in the current formation, which
produces the next prominent current steps in the IV-characteristics.
Consequently, the differential conductance $G=dI/dV\propto d\nu (E)/dE$ of
the device plotted in Fig. \ref{gv}(d) exhibits several pronounced peaks,
the 'main' one at $V_{S}=9.8$mV followed by two at $14.6$ and $15.5$ mV,
each characterising the energetic width and transparency of a resonant
impurity state, whereas in between, at $V_{S}\leq V\leq V_{H1}$, the
differential conductance displays the derivative of the LDOS with respect to
energy.

The energy dependence of the LDOS is the result of the energy-dependent
quantum interference pattern for quasi-particles in the emitter at the
coordinate of a spectrometer, see Fig. \ref{gv}(c). This pattern is random
and tends to reflect an individual portrait of a disordered potential in a
metal surrounding the spectrometer. In the sample under investigation, such
a pattern can be analyzed within the energy interval of about $0\leq E\leq 2$
meV below the Fermi level, since the voltage and energy scales are related
via $E=\alpha e(V-V_{S}),\alpha \approx 0.5$. The correlation energy, $%
E_{c}=\Gamma +\hbar \gamma $, of the fine structure in the differential
conductance pattern is determined by either the energetic spectrometer
width, $\Gamma $, or by the inelastic broadening of states in the emitter, $%
\hbar \gamma $, whichever is larger. According to the theory \cite{fal97},
one can relate $E_{c}$ to the amplitude (rms value) of the fluctuation
pattern of $G(V)\propto d\nu (E)/dE$. Note that, although oscillations at
larger energy scales are also present in each realization of $\nu (E)$,
their contribution to $G(V)$ is suppressed, due to the differentiation. For
a given sample, the spectrometer width, $\Gamma $, can be extracted from the
width of the 'main' resonance peak. For the peak at $V_{S}=9.8$mV in Fig. 
\ref{gv}(d), we find $\Gamma \approx e\alpha \times 72\mu $V $\approx 36\mu $%
eV. For a broad spectrometer, with $\Gamma \gg \hbar \gamma (E)$ at any
excitation energy \cite{sch96}, both the amplitude and correlation voltage
(energy) of fluctuations would be the same over the entire range of $%
V_{S}\leq V\leq V_{H1}$. For a narrow spectrometer, such as studied in the
present work, inelastic broadening of states in the bulk exceeds the
spectrometer width upon increasing the excitation energy of a quasi-hole
left in the emitter. This results in a decrease of fluctuations upon
increasing voltage, as indicated by the dashed lines in Fig. \ref{gv}(d).

Quantitative information about the quasiparticle decay rate, $\gamma (E)$,
is obtained from statistical analysis of the complete fluctuation pattern $%
G(V,B)=dI/dV$studied as a function of a magnetic field, $B$ (applied
parallel to the current flow). Figure \ref{cult} shows a color-scale image
of the differential conductance measured as a function of bias voltage and
magnetic field (in the region of low magnetic fields, where Landau
quantization is hindered by disorder). Sharp black lines in Fig. \ref{cult}
correspond to the spectrometer crossing the emitter Fermi level. The
decrease of the amplitude of observed LDOS fluctuations and the increase of
the correlation voltage of the fluctuation pattern, as a function of the
quasi-hole excitation energy is apparent from the change in the color and
contrast of this image.

As a quantitative measure of the fluctuation amplitude, we calculate the
variance var$_{B}G=\langle \delta G^{2}(B)\rangle _{B}$ using $\delta
G(B)=G(B)-\langle G\rangle _{B}$, where $\langle ...\rangle _{B}$ indicates
averaging over magnetic field in the range of $0\leq B\leq 1.0$ T. Figure 
\ref{result}(a) shows that var$_{B}G$ drops by more than one order of
magnitude within the experimentally accessible voltage range. In our limit
of classical magnetic fields, the fluctuation amplitude is related to the
relaxation rate \cite{fal97} according to 
\begin{equation}
\text{var}_{B}G|_{V}=G_{N}^{2}\times \lbrack 1+\hbar \gamma (E)/\Gamma
]^{-3/2}\;\;.  \label{varg}
\end{equation}
Here, $G_{N}$ is a prefactor which we determine as $G_{N}^{2}=$ var$%
_{B}G_{|V_{S}}$ from Fig. \ref{result}(a) by assuming that, at $V=V_{S}$
(corresponding to $E=0$), $\hbar \gamma \ll \Gamma $ and $E=\alpha e(V-V_{S})
$ is the excitation energy of a quasi-hole (we remind that $\alpha \approx
0.5$ in this experiment).

Figure \ref{result}(b) shows the obtained energy dependence of the
quasi-particle decay rate. It drops strongly upon decreasing the excitation
energy, and, in contrast to some experiments measuring dephasing rates \cite
{moh97}, we do not observe a saturation of $\gamma (E)$ at low energies. In
Fig. \ref{result}(b) we also compare the experimentally determined
quasi-particle relaxation rate with the values calculated using the
Altshuler-Aronov theory of electron-electron interaction in disordered
conductors \cite{alt85,siv94}. It is worth mentioning that the
above-presented determination of the quasi-particle decay rate based upon
the LDOS pattern analysis enables us to study quasi-particles with pretty
small excitation energies, $E<\hbar /\tau $ and to detect the features of
their inelastic decay specific to strongly disordered systems. This make it
different from the analysis of the same quantity on the basis of
measurements of Landau level broadening \cite{mai00}, which requires
distinct Landau quantization and a strong magnetic field (or absence of
impurities).

Electron-electron scattering with a large momentum transfer between
ballistic quasi-particles results in a rate which is determined by the phase
volume of available final states, thus leading to $\gamma \propto E^{2}$ 
\cite{pin66}. In disordered Fermi liquids, where transport is diffusive,
small momentum transfers play an important role, such that an additional $%
E^{3/2}$ energy dependence of $\gamma $ appears, as described in Eq. (\ref
{eqgamma}). The $E^{3/2}$ dependence dominates at small energies, $E<\hbar
/\tau $, while the $E^{2}$ dependence is specific to large energies $E\gg
\hbar /\tau $. After estimating the elastic scattering time of $\tau =0.14$%
ps from the magnetic-field dependence of the LDOS fluctuations \cite{tau}
(and also from the nominal doping level of the emitter contact), we find
that the latter crossover would occur at $E\sim 5$ meV, which is beyond the
energy range accessible in the reported experiment. For a quantitative
comparison, we fit the relaxation rate as $\gamma =b\times E^{2}+A\times
E^{3/2}$ (dashed lines in Fig. \ref{result}(b) show separately the $b\times
E^{2}$ and $A\times E^{3/2}$ parts determined in this fit). We have also fit
the data to $\gamma (E)=A\times E^{x}$ dependence, treating exponent $x$ as
a free parameter, which yields $x=1.54\approx 3/2$ and $A=8\times 10^{10}$%
~meV$^{-3/2}$s$^{-1}$. In the theory \cite{siv94}, the prefactor $A$ in the $%
E^{3/2}$ dependence is $A=(105\sqrt{3\hbar })/(16\pi \tau ^{3/2}E_{F}^{2})$.
After estimating $E_{F}=26$meV from the electron density in the emitter, we
evaluate $A\approx 1\times 10^{11}$~meV$^{-3/2}$s$^{-1}$ which compares well
with the experimental result.

In conclusion, we presented a measurement of the inelastic quasi-particle
relaxation rate in a disordered Fermi liquid. This quantity was obtained
from the analysis of the magnitude of disorder-induced fluctuations in the
local density of states probed using the method of resonant tunneling
through an impurity state. Quantitative comparison with the standard theory 
\cite{alt85} shows that, within the energy range available for such an
analysis, the experimentally determined values of the inelastic relaxation
rate can be attributed to the electron-electron interaction relaxation
mechanism in diffusive conductors.

We thank A. F\"{o}rster and H. L\"{u}th for growing the double-barrier
heterostructure. We acknowledge financial support from BMBF, DFG, EPSRC,
NATO, and TMR.

\begin{figure}[tbp]
\caption{A typical inelastic relaxation process:
a quasi-hole ($\circ $) at energy $E$ below the Fermi level $E_{F}$
(a) decays into a quasi-particle ($\bullet $) and two quasi-holes (b). }
\label{sketch}
\end{figure}

\begin{figure}[tbp]
\caption{(a) Schematic of a strongly asymmetric double-barrier
heterostructure containing residual impurities in the quantum well. (b)
Scanning electron micrograph of the fabricated mesa. The disk on top of the
mesa is a metallization which serves as Ohmic contact. (c) Conduction-band
profile of the device under operation. (d) Differential tunneling
conductance $G=dI/dV$ numerically obtained from $I(V)$ data recorded at the
base temperature of $T=20$ mK of our dilution refrigerator.}
\label{gv}
\end{figure}

\begin{figure}[tbp]
\caption{Color map of the differential conductance as a function of bias
voltage (step 7~$\protect\mu$V) and magnetic field $B\parallel I$
(step~10~mT) for $T=20$~mK, which indicates that the magnitude of 
the fluctuations decreases with increasing bias.}
\label{cult}
\end{figure}

\begin{figure}[tbp]
\caption{(a) Variance of the differential conductance vs bias voltage. (b)
Quasi-particle relaxation rate plotted as a function of excitation energy
measured from the Fermi level. The experimental data are compared with a fit
according to theory for disordered Fermi liquids, $\protect\gamma
=bE^{2}+AE^{3/2}$ (solid). The individual
terms $bE^{2}$ (dash-dotted) and $AE^{3/2}$ (dashed) are
shown as well.}
\label{result}
\end{figure}


\begin{references}
\bibitem{tau}  The elastic scattering time $\tau $ has been obtained from
the measured magnetic field dependence of the variance at high magnetic
fields using var$G/$var$G|_{B=0}=1+(\omega _{c}\tau )^{2}$, where $\omega
_{c}=eB/m^{\ast }$ is the cyclotron frequency~\cite{fal97}.
\end{references}
\end{document}